\documentclass[prb,preprint]{revtex4-2} 
\usepackage{amsmath}  % needed for \tfrac, \bmatrix, etc.
\usepackage{amsfonts} % needed for bold Greek, Fraktur, and blackboard bold
\usepackage{graphicx} % needed for figures
\usepackage{enumitem}

\def\ben{\begin{equation}}
\def\een{\end{equation}}

  \let\n=\nu

\let\C=\Chi

 \def\bd{\begin{document}} \def\ed{\end{document}}
\def\ds{\documentstyle} \let\fr=\frac \let\bl=\bigl \let\br=\bigr
\let\Br=\Bigr \let\Bl=\Bigl
\let\bm=\bibitem
\let\na=\nabla
\let\pa=\partial \let\ov=\overline
\newcommand{\be}{\begin{equation}}
\newcommand{\ee}{\end{equation}}
\def\ba{\begin{array}}
\def\ea{\end{array}}
\def\ft#1#2{{\textstyle{{\scriptstyle #1}\over {\scriptstyle #2}}}}
\def\fft#1#2{{#1 \over #2}}
\def\del{\partial}
\def\vp{\varphi}
\def\sst#1{{\scriptscriptstyle #1}}
\def\oneone{\rlap 1\mkern4mu{\rm l}}
\def\td{\tilde}
\def\wtd{\widetilde}
\def\ie{\rm i.e.\ }
\def\dalemb#1#2{{\vbox{\hrule height .#2pt
        \hbox{\vrule width.#2pt height#1pt \kern#1pt
                \vrule width.#2pt}
        \hrule height.#2pt}}}
\def\square{\mathord{\dalemb{6.8}{7}\hbox{\hskip1pt}}}
\newcommand{\ho}[1]{$\, ^{#1}$}
\newcommand{\hoch}[1]{$\, ^{#1}$}
\newcommand{\bea}{\begin{eqnarray}}
\newcommand{\eea}{\end{eqnarray}}
\newcommand{\ra}{\rightarrow}
\newcommand{\lra}{\longrightarrow}
\newcommand{\Lra}{\Leftrightarrow}
\newcommand{\bp}{\tilde \beta^\prime}
\newcommand{\tr}{{\rm tr} }
\newcommand{\Tr}{{\rm Tr} }
\def\0{{\sst{(0)}}}
\def\1{{\sst{(1)}}}
\def\2{{\sst{(2)}}}
\def\3{{\sst{(3)}}}
\def\4{{\sst{(4)}}}
\def\5{{\sst{(5)}}}
\def\6{{\sst{(6)}}}
\def\7{{\sst{(7)}}}
\def\8{{\sst{(8)}}}
\def\n{{\sst{(n)}}}
\def\cA{{{\cal A}}}
\def\cB{{{\cal B}}}
\def\cF{{{\cal F}}}
\def\cH{{{\cal H}}}
\def\tV{\widetilde V}
\def\tW{\widetilde W}
\def\tH{\widetilde H}
\def\tE{\widetilde E}
\def\tF{\widetilde F}
\def\tA{\widetilde A}
\def\im{{i}}
\def\tY{{{\wtd Y}}}
\def\ep{{\epsilon}}
\def\vep{{\varepsilon}}
\def\R{\rlap{\rm I}\mkern3mu{\rm R}}
\def\bD{{{\bar D}}}

\def\R{\rlap{\rm I}\mkern3mu{\rm R}}
\def\bD{{{\bar D}}}
\def\R{{{\Bbb R}}}
\def\C{{{\Bbb C}}}
\def\H{{{\Bbb H}}}
\def\CP{{{\Bbb C}{\Bbb P}}}
\def\RP{{{\Bbb R}{\Bbb P}}}
\def\Z{{{\Bbb Z}}}
\def\bA{{{\Bbb A}}}
\def\bB{{{\Bbb B}}}
\def\bC{{{\Bbb C}}}
\def\bD{{{\Bbb D}}}
\def\bE{{{\Bbb E}}}
\def\bZ{{{\Bbb Z}}}
\def\Re{{{\frak{Re}}}}
\def\Im{{{\frak{Im}}}}
\def\cosec{{\,\hbox{cosec}\,}}
\def\Gm{{\Gamma_{\!\! -}}}
\def\Gp{{\Gamma_{\!\! +}}}
\def\stan{{standard }}
\def\nonstan{{supernumerary }}

\newcommand{\auth}{Zhiwei Chong}
\begin{document}

% Be sure to use the \title, \author, \affiliation, and \abstract macros
% to format your title page.  Don't use lower-level macros to  manually
% adjust the fonts and centering.

%\title{A Close Look at Inelastic Collisions}
% In a long title you can use \\ to force a line break at a certain location.
\vspace{10pt}

\begin{center}

{\large {\bf A Close Examination on Inelastic Collision}}\\
%\vspace{5pt}
%\auth\\
\vspace{10pt}{}
%\footnote{chong.zhiwei@yahoo.com} 
%\it International Division, Experimental School Affiliated with Zhuhai No.1 High School, Zhuhai, Guangdong, China}\\

%\vspace{10pt}{ \footnote{runnerwei@qq.com} \it Zhuhai No.1 High School, Zhuhai, Guangdong, China}

\end{center}

%When submitting the manuscript for review, do not include the author's name or institution
%\author{Daniel V. Schroeder}
%\email{dschroeder@weber.edu} % optional
%\altaffiliation[permanent address: ]{101 Main Street, Anytown, USA} % optional second address
% If there were a second author at the same address, we would put another 
% \author{} statement here.  Don't combine multiple authors in a single
% \author statement.
%\affiliation{Department of Physics, Weber State University, Ogden, UT 84408-2508}
% Please provide a full mailing address here.

%\author{David P. Jackson}
%\email{ajp@dickinson.edu}
%\affiliation{Department of Physics, Dickinson College, Carlisle, PA 17013}

% See the REVTeX documentation for more examples of author and affiliation lists.

\author{Zhiwei Chong}
\email{chong.zhiwei@gmail.com}
%\altaffiliation[permanent address: ]{101 Main Street, Anytown, USA} % optional second address
% If there were a second author at the same address, we would put another 
% \author{} statement here.  Don't combine multiple authors in a single
% \author statement.
\affiliation{International Division, Experimental School Affiliated with Zhuhai No.1 High School}
\date{\today}
\begin{abstract}
This paper examines the details of an inelastic collision when a bullet shoots a block vertically upward from below. 
With the assumption of constant interaction force between them, we obtain quantities of interest including the displacement for the block at the end of collision, the collision time, and in particular, the bullet's depth inside the block, which are not studied under the traditional assumption of very short collision time. 
We identify the condition under which the collision time can indeed be considered as negligibly short. 
The theory is applied to a well-known demonstration. 
Using the data extracted from photos of the demonstration, we calculate the magnitude of interaction force and the collision time, which can in no way be easily measured otherwise.
\end{abstract}
% AJP requires an abstract for all regular article submissions.
% Abstracts are optional for submissions to the "Notes and Discussions" section.

\maketitle % title page is now complete

\newpage

\section{Introduction} % Section titles are automatically converted to all-caps.
% Section numbering is automatic.
%\section{}
%Ballistic pendulum is a standard problem in almost all introductory physics textbooks. 
%It is used to measure the speed of a bullet by measuring the height a targeted block swings up. 
%The standard way to solve this problem is to apply the conservation of momentum to get the speed of the bullet-block system \emph{after}  the bullet embeds itself in the block. Then the total kinetic energy of the system is transformed to the gravitational potential energy at the highest point. The potential difference is calculated with respect to the block's position before it was hit by the bullet. In other words, an implicit assumption
Some problems in college physics are treated in an over-simplified manner. 
One such simplification is the assumption of very short collision time in an inelastic collision when finding the maximal height of a block shot by a bullet from below in textbook~\cite{giancoli} and the ``thought-provoking" demonstration in~\cite{thought-provoking, another-surprise, veritasium}. 
This approach, which we call the standard approach, is divided into two stages. 
The first stage is the collision between the bullet and the block in which the total momentum of the system is regarded to be conserved even though the net external force on the system, the gravity, is obviously nonzero.
In the second stage, both move up together, and the maximal height is determined by the law of mechanical energy conservation.
Other examples with similar over-simplification include the ballistic pendulum \cite{sears, kleppner, ballistic, slow} and billiard collisions~\cite{frontal, groove}. 

The standard approach needs to be examined closely for the following reasons. 
Firstly, it does not consider the bullet's depth inside the block, which is strikingly exhibited in Fig.~4 of Ref.~\cite{thought-provoking}. 
Secondly, it is not self-consistent; mechanical energy is lost but there is no penetration for the bullet in the block since it treats their vertical displacements as the same. In reality, the bullet's vertical displacement is greater than the block's, and the loss in mechanical energy is due to the negative work done by the force between the bullet and the block.
%the same change in the vertical displacement for the two objects in the standard approach defies the bullet's depth inside the block, 
%which contradicts with the loss in mechanical energy.
Lastly, there arises a puzzle.
The block is considered to remain stationary during the process when the bullet embeds itself in the block, meanwhile it acquires a finite velocity during exactly the same process. 
In other words, it is reasonable to think that the block starts moving at the very moment the bullet contacts the block. 

It is worth pointing out that the introduction of the assumption of very short collision time is due to the lack of information about the complicated bullet-block interaction. 
This is the reason that the standard approach is silent on the bullet's depth inside the block even though it admits the loss in mechanical energy.
Therefore, to study the bullet's depth inside the block, it is necessary to make an assumption on the bullet-block interaction.
As a first step, resorting to Occam’s razor for the moment \cite{Occam}, we make the simplest possible assumption: constant bullet-block interaction.
%As a first step, this paper makes the simplest possible assumption: constant interaction force.

The aim of this paper is to clarify the condition under which the collision time can indeed be considered as short and explain the reason that the block's displacement can be neglected during collision. 
%In other words, it is to find the block's displacement at the end of collision when the collision time is not negligibly short anymore.
Moreover, the bullet's conspicuous depth inside the block exhibited in the impressive demonstration in Ref.~\cite{thought-provoking} provides an excellent opportunity to check the theory against experiment.
%Moreover, it checks the theory against the experimental observations in~\cite{thought-provoking}.

%To address these issues, this paper introduces a constant interaction force between the bullet and the block during collision, which is the minimal step beyond the assumption of very short collision time. 

Exactly the same issue as in this paper is addressed to the ballistic pendulum problem in Ref.~\cite{slow}. 
However, a rather complicated differential equation is derived and numerical algorithm has to be invoked. 
The simple setup of the problem in this paper has two advantages.
It leads to conceptually and analytically transparent results without losing any physics of interest.
More importantly, the theoretical results can be directly used to analyze experimental observations in Ref. \cite{thought-provoking}.
In particular, according to the data that can be extracted from the photos in Ref.~\cite{thought-provoking}, the magnitude of the interaction force during collision and the collision time can be calculated, which can in no way be directly measured otherwise.

%This paper closely examines the inelastic collision followed by mechanical energy conservation, aiming to clarify the long-standing confusion in textbooks and among students.

%Section \ref{standard} presents the results in the standard approach for the purpose of comparison. In Section \ref{constant}, we have the following findings for constant interaction force during collision in comparison with those in the standard approach. 
%Finally, we pin down the condition on the force so that the collision time can be considered as very short. We conclude and discuss in Section~\ref{con}. 

\section{The Standard Approach: very short collision time}\label{standard}
The statement of the problem  or the setup of the experiment is as follows \cite{giancoli, thought-provoking}.
A block of mass $M$ lies on a horizontal table. 
There is a small hole beneath it through which a bullet is shot vertically up at the center of the block. The mass of the bullet is $m$ and its speed when hitting the block is $u$. 
The quantities of interest include the common speed they move up together, the time to reach the highest point, the maximal vertical displacement, and the loss in mechanical energy.
%The standard approach is to divide the process into two parts: 1) the bullet collides with the block inelastically achieving a common speed 
%$v_0$ by momentum conservation, 2) the system consisting of the bullet and the block moves vertically upward with a common velocity, and the height they can arrive is determined by mechanical energy conservation. Meanwhile, it is emphasized that momentum conservation can still be applied because the collision time is so short that the impulse from the force of gravity can safely be ignored.  

The solution is divided into two stages. In the first stage, based on the assumption of very short collision time, the impulse from gravity is ignored. Thereby, the total momentum is \emph{approximately} conserved, that is,
\bea
m\,u\,=\,(M\,+\,m)\,v,
\label{conservation}
\eea
where $v$ is the common velocity of the bullet and the block at the end of collision. It is obtained as 
\bea
v\,=\,\fft{m}{M\,+\,m}\,u. \label{v}
\eea
Then, both move up with the initial speed~$v$ and constant deceleration~$g$ due to gravity. 
%Due to very short collision time, the impulse from the external force, gravity, is non-vanishing but negligible.
%The impulse from the external force, gravity, is actually non-vanishing. However, due to very short collision time, it is negligible.
%In this sense, the momentum is approximately conserved.
%No particular attention is paid to the displacement covered by the block during collision, and it is usually considered to be negligible due to the assumption of very short collision time, which clearly lacks a convincing explanation. Nevertheless, we continue to calculate the maximal height the block can reach.

In the second stage, by the law of mechanical energy conservation or simply by kinematic formulas, the maximal height~$h$ for the block is determined by
\bea
\fft12\,(M\,+\,m)\,v^2\,=\,(M\,+\,m)\,g\,h, \quad 
h\,=\,\fft{v^2}{2\,g}.\label{h}
%\,=\,\left ( \fft{m}{M\,+\,m}\right )^2\,\fft{u^2}{2\,g}.
\eea
It is worth pointing out that the bullet and the block have the same vertical displacement in this approach, which implies that the bullet does not penetrate into the block, or the penetration is negligible.

The time $t$ to reach the highest point is 
\bea
t\,=\,\fft{v}{g}.\label{t}
\eea
The loss in mechanical energy $L$ is the difference between the bullet's initial and the system's total kinetic energy at the end of collision, that is,
\bea
L\,=\,\fft12\,m\,u^2\,-\,\fft12\,(M\,+\,m)v^2
   \,=\,\fft{M}{M\,+\,m}\,\fft12\,m\,u^2,\label{L}
\eea
which is a fraction of the bullet's kinetic energy just before collision.
The quantities $v, h, t$, and~$L$ in Eqs. (\ref{v}), (\ref{h}), (\ref{t}), and (\ref{L}), respectively,  will be compared with their counterparts in Sec. \ref{constant}.

A few remarks are in order here. The assumption of very short collision time plays a crucial role in this approach. 
On the one hand, the impulse from gravity, which is the product of gravity and the collision time, can safely be ignored due to this assumption.
This is the very reason that we can apply the law of momentum conservation approximately in Eq.~(\ref{conservation}). 
%Thereby, the momentum of the system is approximately conserved as in Eq. (\ref{v}).
On the other hand, it helps to explain why there is a finite velocity but negligible displacement at the end of collision.
The velocity of the block is the product of the \textit{average} acceleration (even though it is not explicitly considered) and the collision time, while its displacement is one half of the product of the former and the \textit{square} of the latter. 
Numerically, when the collision time is very short, the block's displacement is a second order infinitesimal, while its velocity is a first order one. Hence, the former is  negligible compared with the latter. 
%If students happen to have some basic knowledge about infinitesimals in calculus, then this does help to clear their doubts. 

%It is normally argued that the impulse from gravity may be neglected due to very short interaction time. However, it is exactly this neglect that is the focus of this paper.
%; we want to clarify the condition under which the interaction may be indeed very short. The other side of the question is that what we shall do if the interaction time is not extremely short. 

%Clearly, we are interested in the consequences when the collision time cannot be considered as very short anymore. 
%3) How to solve the problem rigorously when this assumption does not hold?
%In other words, we want to go beyond the assumption of negligibly short collision time, which is the focus of the next section.

\section{Constant Interaction Force}\label{constant}
%Strictly speaking, the momentum is not conserved since there is non-vanishing external force gravity acting on the system during collision. 
%In other words, the momentum is at most approximately conserved. 
%On the other hand, the block starts moving upward at the very moment the bullet hits it. 
%Moreover, the displacements for the bullet and the block should \emph{not} be the same; the displacement for the bullet is larger than that for the block since the bullet embeds itself inside the block.
%[We will see that neglecting collision time is actually equivalent to neglecting the depth the bullet embeds itself inside the block.]

The reason for introducing the assumption of negligible collision time is the lack of knowledge about the interaction force. 
Thereby, to go beyond the assumption of very short collision time, we need a plausible assumption on the bullet-block interaction: constant interaction force.
 
%The focus is on quantities like 
%the collision time, 
%the depth the bullet embeds itself in the block, 
%the displacement the block covers during collision, 
%the total time to reach the highest point, 
%the maximal height of the block, 
%the maximal height of the center of mass of the system,
%and certainly the comparison of these quantities with those in the standard approach obtained in Sec. \ref{standard}.
%To delve into such details of the collision process, we need details about the interaction force during collision. In this section, we start with the simplest possible assumption on the force between the bullet and the block, that is, constant force $k_0$ during the whole process of collision.
%The subscript 0 results from the fact that constant interaction force can be considered as depending on the zero-th power of the depth that the bullet embeds inside the block. We will consider forces depending on non-zero powers of depth in a future work. 
With this assumption, more details of the collision can be studied. 
In particular, we can calculate quantities such as the collision time, the bullet's depth inside the block, and the block's vertical displacement at the end of collision, which are not addressed in the standard approach. 
%Moreover, it is also of interest to compare with those quantities obtained in Sec.~\ref{standard}.
%between the two approaches such quantities like the common speed at which they start moving up together, the maximal displacement of the block in the vertical direction, the total time to reach the highest point, the loss in mechanical energy, and the maximal height for the center of mass. 
In addition, we pin down the condition under which the collision time can be considered as very short: the interaction force is very large compared with the block's weight. 
Lastly, all the results under constant but very large interaction force reduce to those in the standard approach.
%The bullet starts embedding itself in the block. Clearly, there is a resistive force acted on the bullet from the block. At the same time, according the Newton's third law, there is a reacting force of the same magnitude acted on the block from the bullet in the opposite direction. For simplicity, we assume the force to be a constant $f$ during the whole embedding process. The acceleration of gravity is denoted as $g$.

%The condition on applying momentum conservation is that there is no net external force acting on the system under consideration. Clearly, the gravity is the non-zero external force in this problem. Strictly speaking, we cannot apply momentum conservation in this case. In other words, we can at most apply momentum conservation in an approximate sense. 

The constant interaction force, the block's acceleration, and the bullet's deceleration are denoted as $F, A$, and  $a$, respectively. 
Applying Newton's second law to each object gives 
\bea
F\,-\,M\,g\,&=&\,M\,A, \quad\quad A\,=\,\fft{F}{M}\,-\,g,\label{A}\\
F\,+\,m\,g\,&=&\,m\,a,\quad\quad a\,=\,\fft{F}{m}\,+\,g. \label{a}
\eea

The collision time $\tau$ or the time when the bullet stops moving inside the block is determined by equating the velocity of the block, $A\,\tau$, with that for the bullet, $u\,-\,a\,\tau$, that~is,
\bea
%A\,\tau\,=\,u\,-\,a\,\tau, \quad 
	\tau\,=\,\fft{u}{A\,+\,a}\,=\,\fft{M}{F}\,\fft{m}{M\,+\,m}\,u
        \,=\,\fft{M}{F}\,v
				\,=\,\fft{v}{r\,g},\quad 
				r\,\equiv\,\fft{F}{M\,g},\label{tau_0}
\eea 
where the second equality follows from Eqs. (\ref{A}) and (\ref{a}), and the third from Eq.~(\ref{v}).
%The quantity $\tau$ is just the collision time. 
The ratio~$r$ plays an important role. 
When it is very large, or equivalently, the force $F$ is very large compared with the the block's weight $M\,g$, the collision time $\tau$ can be considered as very short. 
%Thereby, $r>>1$ is the condition under which the collision time can indeed be considered as negligibly short.
Moreover, we will soon see that in the large~$r$ limit, all the results do reproduce those in Sec.~\ref{standard}.

Under the assumption of constant interaction force, we are able to calculate the block's vertical displacement $y'$ at the end of collision, which is not studied in the standard approach. Applying the kinematic formula gives
\bea
y'\,=\,\fft12\,A\,\tau^2
 \,=\,\fft12\,(r\,-\,1)\,g\,\left(  \fft{v}{r\,g}\right)^2
 \,=\,\fft{r\,-\,1}{r^2}\,\fft{v^2}{2\,g}
 \,=\,\fft{r\,-\,1}{r^2}\,h, 
\label{y'}
\eea
where the second equality follows from $A$ in Eq. (\ref{A}) and $r$ and $\tau$ in Eq. (\ref{tau_0}).
The last equality follows from $h$ in Eq. (\ref{h}). As expected, when $r$ is very large, $y'$ goes to zero. This is the reason that the block's vertical displacement is considered to be negligible in the standard approach. 
%Thereby, $r\rightarrow\infty$ is the condition under which the displacement for the block at the end of collision can be ignored.
%A remark is in order here. In the standard approach, the collision time is assumed to be very short so that there is no displacement for the block at the end of collision. In reality, there is a displacement covered by the block no matter how short the collision time $\tau_0$ is. 

The common velocity $v'$ at which the block and the bullet start moving up together is
\bea
v'\,=\,A\,\tau\,=\,(r\,-\,1)\,g\,\fft{v}{r\,g}\,=\,\left( 1\,-\,\fft{1}{r} \right)\,v.\label{v_0}
\eea
We see that $v'<v$, and $v'$ goes to $v$ in Eq. (\ref{v}) when $r$ goes to $\infty$.

A remark is in order here. We see that the common velocity $v'\,=\,A\,\tau$ is proportional to the collision time $\tau$, meanwhile the block's displacement $y'\,=\,\fft12\,A\,\tau^2$ at the end of collision is proportional to $\tau^2$. 
If the collision time $\tau$ is very short, then the displacement for the block $y'$ is negligible compared with its velocity $v'$. 
Mathematically speaking, the common velocity $v'$ is a first order infinitesimal in $\tau$ and the displacement for the block $y'$ is a second order infinitesimal. 
This is the very mathematical reason that the block acquires a finite velocity while its displacement is negligible at the end of collision under the assumption of very short collision time. 
%We can make the above observation even more precise. The ratio between $h_0$ and $v_0$ is
%\bea \fft{X_0(\tau_0)}{v_0}\,=\,\fft{1}{r}\,\fft{h}{v}. \eea
%We see that in the large $r$ limit, the displacement for the block $X_0(\tau_0)$ at the end of collision is negligible compared with its $v_0$.
%For students with a bit background in calculus, this observation would help them understand the viability of the standard approach. Unfortunately, students who take this level of physics normally do not have such luxury in mathematics. However, this won't prevent us from notifying the students in advance that they will understand this point someday in the future.

%The quantity $H_0$ is actually the unit of length for this problem and $v_0$

The block's maximal vertical displacement $h'$ is the sum of its displacement~$y'$ at the end of  collision and that covered in the upward projectile motion thereafter, that is,
\bea
h'\,=\,y'\,+\,\fft{v'^2}{2\,g}
  \,=\,\fft{r\,-\,1}{r^2}\,h\,+\,\left( 1\,-\,\fft{1}{r} \right)^2\,\fft{v^2}{2\,g}
	\,=\, \left( 1\,-\,\fft{1}{r} \right)\,h,\label{h_0}
\eea
where the expression for $h$ in Eq. (\ref{h}) is used to obtain the last equality.
We see that $h'\,<\,h$, and $h'\,\to\,h$ as $r\,\to\,\infty$.
Figure 1 compares  between the two approaches their relative positions for the block and the bullet at their highest points. 
\begin{figure}[h!]
\centering
\includegraphics[width=4.0in]{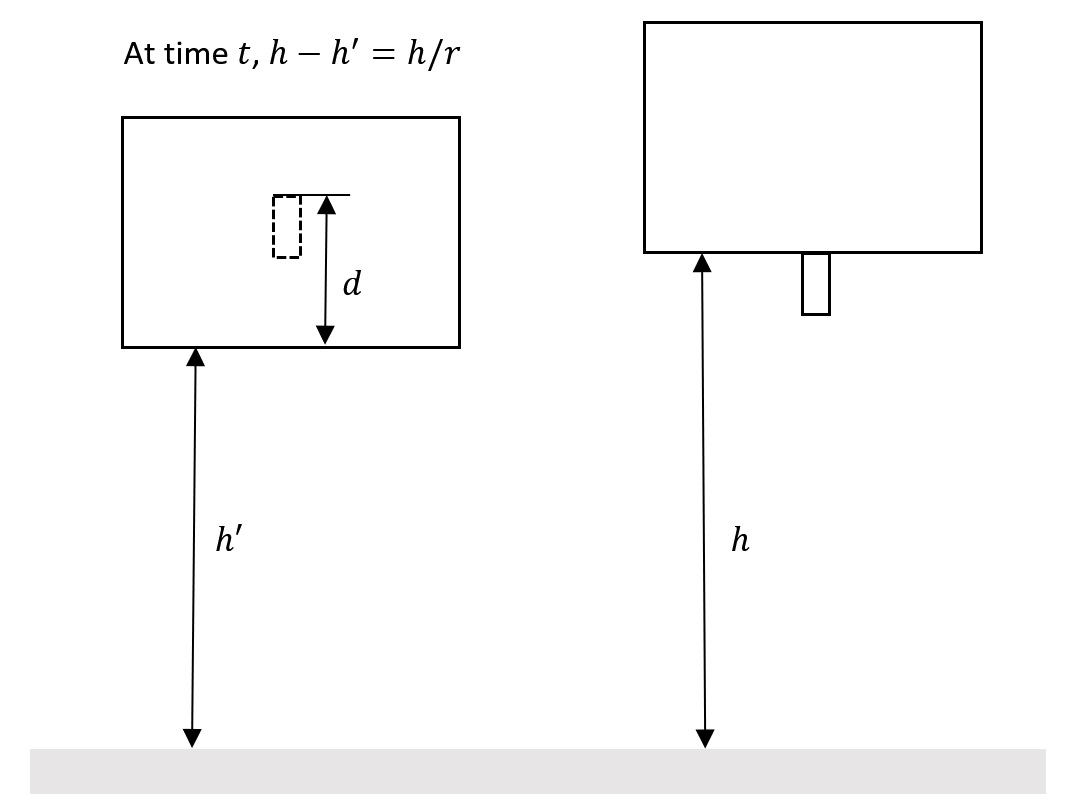}
%\hspace{1cm}。jpg
%\label{setup}
\caption{Comparison between two approaches at their maximal displacements. The left figure shows the situation under the assumption of constant interaction force, and the right is for the standard approach.}
\centering
\end{figure}

The total time for the block to reach the highest point is the sum of the collision time $\tau$ and that for the upward projectile motion, that is,
\bea
t'\,=\,\tau\,+\,\fft{v'}{g}
  \,=\,\fft{v}{r\,g}\,+\,\left(  1\,-\,\fft{1}{r}\right)\,\fft{v}{g}
	\,=\,\fft{v}{g}
	\,=\,t,
\eea
where the second equality is obtained from  $\tau$ in Eq.~(\ref{tau_0}) and $v'$ in Eq.~(\ref{v_0}), and the last equality follows from Eq. (\ref{t}).
A bit surprisingly, it takes the same amount of time to reach the highest point as in the standard approach. 
We believe it to be a coincidence which might depend on the specific form of interaction force.
%This result does not depend on the magnitude of the constant interaction force $F$.
%In a future work, we will show that this result remains true as long as the interaction force only depends on the depth for the bullet inside the block.

The bullet's maximal depth inside the block is the product of its average speed with respect to the block,~$\fft12\,u$, and the collision time~$\tau$, that is,
\bea
d\,=\,\fft12\,u\,\tau
	\,=\,\fft{1}{r}\,\left( 1\,+\,\fft{M}{m} \right)\,\fft{v^2}{2\,g}  
	\,=\,\left( 1\,+\,\fft{M}{m} \right)\,\fft{h}{r},
\label{d}
\eea
where the second equality follows from Eqs. (\ref{v}) and (\ref{tau_0}), and the third from Eq. (\ref{h}).
The first equality needs more explanation. 
The bullet moves inside the block with constant acceleration. 
Before the collision, its speed with respect to the block is $u$. At the end, it becomes~0. Thereby, the average is ~$\fft12\,u$.
An alternative way to obtain this result is to find the difference in displacement between the bullet and the block at the end of collision, which gives the same answer in Eq. (\ref{d}).

%[more work needed] Equation (\ref{d}) shows that as the collision time is very short, the depth that the bullet penetrates inside the block is negligible. This is an important observation; very short collision time implies negligible depth for the bullet inside the block. In other words, the bullet's striking depth inside block suggests that the assumption of very short collision time is inappropriate for the bullet-block collision in \cite{thought-provoking}.

It is also of interest to compare the bullet's vertical displacement in the two approaches. In the standard approach, it is equal to the block's displacement $h$ in Eq. (\ref{h}). With constant interaction force, it is the sum of the block's displacement $h'$ in Eq. (\ref{h_0}) and the bullet's depth $d$ in Eq. (\ref{d}), that is,
\bea
h'\,+\,d%\,=\,\left( 1\,-\,\fft{1}{r} \right)\,h\,+\,\fft12\,u\,\tau
				\,=\,\left( 1\,-\,\fft{1}{r} \right)\,h\,+\,\left( 1\,+\,\fft{M}{m} \right)\,\fft{h}{r}
				\,=\,\left( 1\,+\,\fft{M}{r\,m} \right)\,h.
\eea
Clearly, it is larger than $h$ in the standard approach.

The loss in mechanical energy $L'$ is the product of the constant force~$F$ and the depth~$d$, that is,
\bea
L'\,=\,F\,d
	\,=\,\fft12\,F\,u\,\tau
	\,=\,\fft12\,F\,u\,\fft{v}{r\,g}
	\,=\,\fft{M}{M\,+\,m}\,\fft12\,m\,u^2,\label{L_0}
\eea
where the last equality follows from $r$ in Eq. (\ref{tau_0}) and $v$ in Eq. (\ref{v}).
The right hand side of Eq. (\ref{L_0}) is exactly the same as that in Eq. (\ref{L}), that is,
\bea
L'\,=\,L.
\label{L'=L}
\eea
It is worth pointing out that $L'$ can also be obtained by calculating the difference between the bullet's kinetic energy just before the collision and the system's gravitational potential energy at the highest point.

The equality in Eq. (\ref{L'=L}) implies that the total increase in gravitational potential energy should be the same in both approaches. Indeed, it can be readily verified~that 
\bea
(M\,+\,m)\,g\,h\,=\,M\,g\,h'\,+\,m\,g\,(h'\,+ \,d), \label{xc0}
\eea
where the left hand side is the increase in gravitational potential energy in the standard approach, and the right is that under constant interaction force. 
Equation~(\ref{xc0}) can alternatively be written as
\bea
h\,=\,\fft{M\,h'\,+\,m\,(h'\,+\,d)}{M\,+\,m},
\eea
where the right hand side is interpreted as the vertical displacement for the center of mass.
%The right hand side is the maximal change of the height for the center of mass for the system. 
In other words, the equality shows that the maximal displacements for the center of mass are the same in both approaches, even though the maximal displacements for the block are different.

%The difference in the maximal height for the block and the bullet can be explained by the same change in the height of the center of mass; the larger height the bullet rises in our approach is compensated by the smaller maximal height of the block to maintain the same maximal height of the center of mass.

The difference in maximal displacement, $h'<h$, can be understood in the following way.
In the standard approach, the depth for the bullet inside the block is neglected, and both climb the same distance in the vertical direction. 
Meanwhile, under constant interaction force, their maximal displacements are not the same anymore; the bullet covers a larger vertical displacement than the block by the amount~$d$ in Eq.~(\ref{d}).
Since the maximal displacements for the center of mass are the same in both approaches, the maximal displacement for the block has to be less than that in the standard approach to compensate for the larger maximal displacement for the bullet.

We summarize our findings as follows.
\begin{enumerate}[nosep]
	\item The total time to reach the highest point, the loss in mechanical energy, and the maximal displacement for the center of mass are the same in both approaches. 
	%\item The loss in mechanical energy is the same for both approaches. 	
	%\item The maximal displacement for the center of mass is the same for both approaches.
	\item The maximal displacement for the block (bullet) under constant interaction force is lower (higher) than that in the standard approach.
%, and the maximal height for the bullet in our approach is higher than that in the standard approach.
	\item In the large $r$ limit, that is, when the force $F$ is very large compared with the block's weight, the results reduce to those in the standard approach. In particular, both the collision time and the bullet's depth inside the block vanish in this limit.
\end{enumerate}

\subsection{Further Comparisons}\label{comparison}
This subsection compares the velocities and displacements for the block at time~$\tau$ in both approaches. 
The velocity at time $\tau$ or at the end of collision under constant interaction force is~$v'$ obtained in Eq.~(\ref{v_0}). 
%In the standard approach, the velocity at the end of collision is $v$ in Eq. (\ref{v}). 
On the other hand, in the standard approach, the velocity for the block at time~$\tau$~is
\bea
v\,-\,g\,\tau\,=\,v\,-\,g\,\fft{v}{r\,g}\,=\,\left( 1\,-\,\fft{1}{r} \right)\,v,
\label{vt}
\eea
which is exactly the same as $v'$ in Eq. (\ref{v_0}).
In other words, the velocities are different at the end of collision for two approaches; it is $v$ in the standard approach and $v'<v$ under constant interaction force. However, they are the same at time~$\tau$.

The height of the block at time $\tau$ or at the end of collision under constant force is $y'$ in Eq. (\ref{y'}).
Meanwhile, the height of the block $y$ in the standard approach at time $\tau$ is
\bea
y\,=\,v\,\tau\,-\fft12\,g\,\tau^2
					\,=\,\fft{1}{r}\,\left( 2\,-\,\fft{1}{r}  \right)\,h.   \label{y}
\eea
The difference between $y'$ in Eq. (\ref{y'}) and $y$ in Eq. (\ref{y}) is
\bea
y'\,-\,y\,=\,\fft{r\,-\,1}{r^2}\,h\,-\,\fft{1}{r}\,\left( 2\,-\,\fft{1}{r}  \right)\,h
				\,=\,-\fft{1}{r}\,h.\label{dy}
\eea
On the other hand, the difference between $h'$ in Eq. (\ref{h_0}) and $h$ in Eq. (\ref{h}) is 
\bea
%\D h_0\,=\,
h'\,-\,h\,=\,\left( 1\,-\,\fft{1}{r} \right)\,h\,-\,h
				\,=\,-\fft{1}{r}\,h.
\eea
%We see that the maximum height $h_0$ the block can arrive under constant interaction force $k_0$ is less than $h$ from the standard approach in Eq. (\ref{h}). 
Thereby,
\bea
h'\,-\,h\,=\,y'\,-\,y.
\eea
In words, the difference in maximal vertical displacement for the block is the same as that at time~$\tau$.
This can be understood in the following way. The velocities at time $\tau$ are the same for the two approaches as shown in Eqs. (\ref{v_0}) and (\ref{vt}). Thereby, the displacement that the block ascends thereafter is the same. As a result, the difference in the maximal vertical displacements must be the same as that at time~$\tau$.
The above findings are illustrated in Fig.~2.
\begin{figure}[h!]
\centering
\includegraphics[width=4.0in]{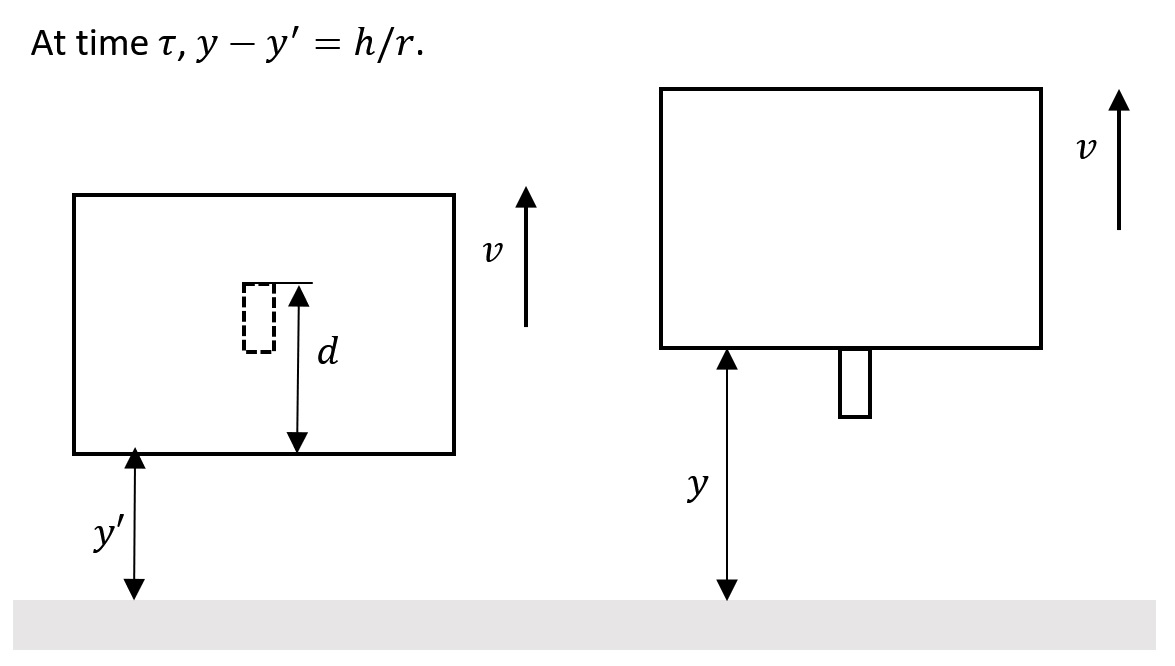}
%\hspace{1cm}。jpg
%\label{setup}
\caption{Comparison between two approaches at time $\tau$. The left figure shows the situation under the assumption of constant interaction force, and the right is for the standard approach.}
\centering
\end{figure}

\section{Estimation of Interaction Force and Collision Time}
%Though the collision time is very short, but the interaction force is very large. 
%The derivation in this paper helps to determine the magnitude of interaction force and the muzzle velocity of the bullet.

The thought-provoking demonstration in Ref. \cite{thought-provoking} provides us with an excellent opportunity to check the theory in this paper. Unfortunately, the emails of the authors are not in use anymore, and we are not able to contact with them to get the data such as the mass of the block and bullet, the bullet's depth inside the block, and the model of the pistol which may provide the bullet's muzzle velocity.
In particular, if we could get the video of the experiment, we would be able to measure the time for the block to reach its maximal height.

The thickness the block is about 7 cm, and the pistol's caliber is~.22 which enables us to estimate the mass and muzzle velocity of the bullet~\cite{thought-provoking}.
Moreover, Figs.~2 and~3 in the reference make it possible to roughly measure the block's maximal height and the bullet's depth inside the block. 
On the upper half of Table.~I, we list the available data as input to our calculation. We will explain how they are obtained in Sec.~\ref{data}.  
%\begin{enumerate}
%	\item The size of the block is 12 by 7 by 7 cm.
%	\item The maximal displacement for the block $h'\approx$ 101.5 cm.
%	\item The bullet's depth inside the block $d\,\approx\, 5.7$ cm.
%	\item The mass of the bullet $m\approx$ 2.268 g.
%	\item The muzzle velocity $u\,\approx\,350.071$ m/s.
%\end{enumerate}
\begin{table}[h!]
  \begin{center}
        
\begingroup

\setlength{\tabcolsep}{8pt} % Default value: 6pt
\renewcommand{\arraystretch}{1.5} % Default value: 1				
    \begin{tabular}{|l|c|c|c|c|c|c|}
      %bullet mass & block & muzzle & block & penetration &collision & interaction \\
			%mass & mass & velocity & height  & depth 	&time	& force \\
			\hline
		Input&	$h'$  & $d$ 	&  $m$ & $u$ &  /    \\
      \hline
		Value &	101.5 cm & 5.7 cm   &  2.268 g & 350.1 m/s  &  / \\
      \hline
			\hline
		Calculated&	$M$ &  $\tau$	& $F$  & $y'$ & $t'$ \\
      \hline
		Value &	  175.68 g &     0.33 ms  &  2394.4 N  & 0.73 mm     & 0.45 s   \\
      \hline
  \end{tabular}
\label{tab:quantities}
\endgroup	
	\caption{Input and Calculated Quantities for the Collision.}
  \end{center}
\end{table}

We are ready to calculate the quantities that are either not available to us or cannot be measured directly.
From Eqs. (\ref{d}), (\ref{tau_0}), and (\ref{v}), the depth can be expressed as
\bea
d\,=\,\fft12\,u\,\tau
	\,=\,\fft12\,u\,\fft{v}{r\,g}
	\,=\,\fft{m}{r\,(M\,+\,m)}\,\fft{u^2}{2\,g}.
	\label{d'}
\eea
On the other hand, the height $h'$ in Eq. (\ref{h_0}) can be expressed in terms of $m$ and $u$ as
\bea
h'\,=\,\left(  1\,-\,\fft{1}{r}  \right)\,h
	\,=\,\left(  1\,-\,\fft{1}{r}  \right)\,\fft{v^2}{2\,g}
	\,=\,\fft{(r\,-\,1)\,m^2}{r\,(M\,+\,m)^2}\,\fft{u^2}{2\,g},\label{h'}
\eea
where the second equality follows from Eq. (\ref{h}) and the third from Eq. (\ref{v}).
Plugging in the values for $m, u, d$, and $h'$ from Table~I into Eqs. (\ref{d'}) and (\ref{h'}) gives
\bea
r\,\approx\,1391,   \quad   M\,\approx\,175.68 g.
\eea
Plugging the values for $r$ and $M$ into Eq. (\ref{tau_0}) gives
\bea
F\,=\,r\,M\,g\,\approx\,2394.42\,\, \mathrm{N},\quad 
\tau\,=\,\fft{v}{r\,g}
		\,=\,\fft{m\,u}{r\,(M\,+\,m)\,g}
		\,\approx\,0.33\,\, \mathrm{ms}.
\eea
Note the magnitude of $F$ is rather large; it is as large as the weight of a mass 244.08~Kg!
The values for $M, \tau $, and $F$ are listed as calculated quantities on the lower half of Table~I. 

Another quantity that is of our interest but cannot be measured directly is the displacement of the block at the end of collision. From Eqs. (\ref{y'}), (\ref{d}), and the data in Table~I, we obtain 
\bea
y'\,=\,\fft{1\,-\,\fft{1}{r}}{1\,+\,\fft{M}{m}}\,d\,\approx\,0.73 \mathrm{mm},
\eea
which is  much smaller than the bullet's depth in the block.

These results are interesting. By measuring the maximal height $h'$ for the block and the bullet's depth $d$ in the block, we are able to calculate the rather large interaction force $F$ on one hand and the very short collision time $\tau$ on the other. These two quantities can in no way be directly measured with any imaginable means.

Finally, the total time $t'$ to reach the highest point is
\bea
t'\,=\,t\,=\,\fft{v}{g}\,=\,\fft{m}{M\,+\,m}\,\fft{u}{g}
	\,\approx\,0.46 \mathrm{s}.
\eea
Had we obtained the video of the demonstration, we could have been able to compare this calculation with the time extracted from video to examine the validity of our assumption.

\subsection{Method to Obtain the Data}\label{data}
The size of the block is given in Ref.~\cite{thought-provoking}, and its thickness is about 7~cm. From Fig.~2 in the reference, the maximal height $h'$ of the block is about 14.5 multiples of the block's thickness, that is, 
$h'\,\approx\,14.5\times 7\,=\,101.5$ cm. From Fig.~4, the bullet's depth inside the block is estimated as $d\,\approx\,5.7 $ cm.

The pistol used in the demonstration is of caliper .22, but the model of the pistol is not provided. However, the mass of this type of bullet and its muzzle velocity can easily be obtained from the internet~\cite{.22}. The mass of the bullet ranges from 30 gr to 40 gr (1~gr~$\approx$~0.065~g). The muzzle velocity ranges from 1070 fps to 1600 fps (1 fps $\approx$ 0.305 m/s). The muzzle energy is between 100 ft.lbf and 105 ft.lbf (1 ft.lbf $\approx$ 1.356 J).
The range for the muzzle energy is rather narrow, and the range for the muzzle velocity is rather wide. Thereby, we use the muzzle energy and the bullet mass to calculate the muzzle velocity.
In the following, the muzzle energy is chosen as 102.5 ft.lbf which is approximately 138.971 J. The mass of the bullet is chosen as 35 gr which is about 2.268 g. 
With the notations in this paper, these values are $\fft12\,m\,u^2\,=\,138.971$ J and $m\,=\,2.268$ g from which the muzzle velocity is obtained as $u\,\approx\,350.1$ m/s.

\section{Conclusion and Discussion}\label{con}
This paper closely examines the details of inelastic collision by introducing the assumption of constant interaction force to replace the assumption of very short collision time. Comparisons are made for various quantities of interest between the two approaches. 
In the large $r$ limit, that is, when the interaction force is very large compared with the weight of the block, then all the results reduce to those under the assumption of very short collision time.

%The assumption of very short collision time comes from the lack of knowledge in the interaction force during collision. By the introduction of constant interaction force, more quantities like the collision time and the depth the bullet embeds itself in the block can be calculated. 
However, there arises one question: to what extent do the results obtained in this paper depend on the peculiarity of the assumption of constant interaction force?   
In other words, will it still take the same amount of time for the block to reach its highest point when the interaction force is not constant anymore? Will the energy loss be the same? 
These are good questions for a student project.
%\begin{acknowledgments}

%We gratefully acknowledge Harvey Gould and Jan Tobochnik, who created an earlier AJP \LaTeX\ sample article that inspired this one.  This work was supported by the American Association of Physics Teachers.

%\end{acknowledgments}

\end{document}